# ADAPTIVE AND SECURE ROUTING PROTOCOL FOR EMERGENCY MOBILE AD HOC NETWORKS

Emmanouil A. Panaousis, Tipu A. Ramrekha, Grant P. Millar and Christos Politis

Wireless Multimedia & Networking (WMN) Research Group
Kingston University London, United Kingdom
{e.panaousis,a.ramrekha,g.millar,c.politis}@kingston.ac.uk

## ABSTRACT

*The nature of Mobile Ad hoc NETworks (MANETs) makes them suitable to be utilized in the context of an extreme emergency for all involved rescue teams. We use the term emergency MANETs (eMANETs) in order to describe next generation IP-based networks, which are deployed in emergency cases such as forest fires and terrorist attacks. The main goal within the realm of eMANETs is to provide emergency workers with intelligent devices such as smart phones and PDAs. This technology allows communication "islets" to be established between the members of the same or different emergency teams (policemen, firemen, paramedics). In this article, we discuss an adaptive and secure routing protocol developed for the purposes of eMANETs. We evaluate the performance of the protocol by comparing it with other widely used routing protocols for MANETs. We finally show that the overhead introduced due to security considerations is affordable to support secure ad-hoc communications among lightweight devices.*

## 1. INTRODUCTION

Mobile Ad hoc Networks (MANETs) consist of a group of mobile and autonomous nodes that are directly or in a multihop manner interconnected to each other (see Figure 1). MANETs do not have any communication infrastructures such as access points or base stations relying entirely on nodes' cooperation for forwarding information from data sources to intended destination nodes. As a result, any wireless mobile devices or nodes participating in the network can act as a source, destination or router.

An emergency critical case is an event, which threatens serious damage to human welfare or to the environment. This situation may be for instance a war or a terrorist attack threatening serious damage to integral parts of a locations infrastructure. In many extreme emergency scenarios such as natural or manmade disasters, the rescuers may face difficulty using traditional legacy networks due to destruction or collapse of the telecommunications infrastructure.

The nature of MANETs makes them suitable to be utilized in the context of an emergency for all the involved emergency rescue teams. Due to their flexibility and self-organization capabilities, MANETs are well suited for scenarios where certain network services such as message routing, event notification and video transmission have to be provided quickly and dynamically without any centralized infrastructure. For instance, there may be situations where emergency workers, illustrated in Figure 2, from different recovery teams (e.g. police, paramedic, fire brigade) must cooperate and coordinate their actions. One major technological challenge is the implementation of a general solution for secure multimedia communications with QoS provisioning for eMANETs. Within this context, recovery workers will be equipped with a personal digital assistant such as a PDA to be able to communicate with each other during a critical rescue mission by exchanging text, voice and video.

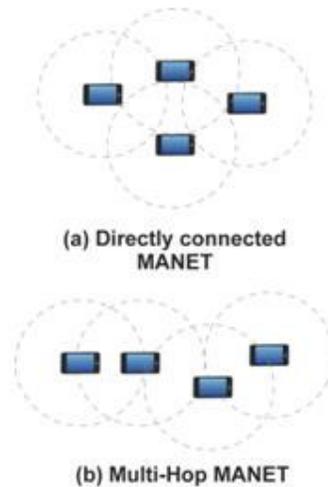

Figure 1. Types of decentralised architectures for MANETs.

Routing in MANETs is crucial due to the lack of a central routing entity. Each intermediate node forms part of the path between a source and a destination acting as a router for the Internet Protocol (IP) datagram. It is worth mentioning at this point that in eMANETs, routing is apparently a more complex problem when compared to the traditional infrastructure based wireless networks. For instance, extreme emergency infrastructures should provide fast, reliable and lightweight secure communication links. Likewise, within the realm of eMANETs the network size will vary whenever more rescuers join or leave the disaster area or critical area (CA) according to the severity of the situation. Battery exhaustion of lightweight mobile communication devices used by rescuers could also stipulate another reason for changes in the network size.

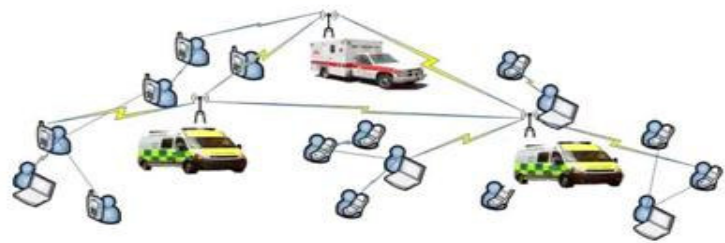

Figure 2. Network architecture for extreme emergency services [21].

Generally speaking, an eMANET routing protocol should operate in a distributed manner and be able to react to both logical and physical changes of the network state. Parameters that define the network state include:

- Node mobility levels
- Node transmission radius
- Battery level of nodes
- Traffic requirements (i.e. priorities)
- Data load

- Wireless link quality
- Network size

To this end, we envisage an architecture that is compatible with existing and future emerging technologies, whilst ensuring fast response time in case of high priority traffic events. This raises the need for candidate routing solutions that can establish a fast and adaptive connection between mobile nodes where the radio link quality is time varying according to the channel propagation characteristics.

Secure operation of the MANET routing protocol is critical because of the absence of a fixed infrastructure. Nodes are associated and will cooperate virtually with any node including adversaries. Adversaries can cause the disruption of the route discovery and data forwarding operations. Disruption of the route discovery can cause systematic problems to the flow of data. In addition, adversaries can obstruct the propagation of legitimate queries and routing updates. In order to prevent such attacks it is important for the receiver node to verify the authenticity of the sender and the integrity of the data.

## 2. FUNDAMENTAL ISSUES

Among the different MANET routing protocols, the MANET Working Group (WG) of Internet Engineering Task Force (IETF) mainly favours the proactive and reactive approaches. In this context, three of the most popular protocols are the reactive Ad-hoc On demand Distance Vector (AODV) [2], the reactive Dynamic Source Routing (DSR) [3] and the proactive Optimized Link State Routing (OLSR) [4] protocols. The proactive OLSR protocol is a variant of traditional link state routing modified for MANETs. The protocol uses a variant of Dijkstra's shortest path algorithm to provide optimal routes considering the number of hops. On the other hand the reactive AODV uses a modified version of the Bellman-Ford algorithm for mobile environments. Finally, DSR requires the sender to know the complete route to destination. It is based on two main processes: (i) the route discovery process which is based on flooding and it is used to dynamically discover new routes and maintain them in node cache and (ii) the route maintenance process during DSR periodically detects and notifies network topology changes.

In MANETs using the OLSR protocol, first described in [4], each node issues periodic "HELLO" and "Topology Control (TC)". The "HELLO" messages are used for new link discovery and for updating existing link information where a link is defined as a connection between two interfaces that are within the same radio coverage area, i.e. they can directly communicate to each other. These messages are sent to one-hop neighbours only. Each node also use network wide TC messages periodically to disseminate topological information that it contains including the routes that it currently maintains. The network wide flooding of TC messages is reduced using Multi-Point Relays (MPR) as explained in [4]. Hence, OLSR readily provides a periodically updated route for data transmission.

AODV is an on-demand routing protocol as described in [2]. In a MANET using AODV routing, a source node has to initiate a route discovery mechanism each time data is ready to be transmitted. The AODV route discovery process consists of flooding the network with a Route Request (RREQ) packet to be received by the intended destination node. The destination node then, sends a unicast Route Reply (RREP) to the source node thus establishing a route between the two nodes for data transmission. This process adds delay towards end-to-end data delivery. However, on the other hand, AODV provides a fresh route for data delivery as opposed to periodically pre-established routes.

Authors in [5] propose a secure version of AODV named SAODV (Secure AODV) using digital signatures, asymmetric encryption keys and hash chains. The protocol provides characteristics such as integrity, non-repudiation of the routing data and authentication of the nodes within a MANET. Actually, SAODV takes advantage of the pure routing functionality of AODV while it adds security mechanisms on top of the conventional AODV protocol. Nodes sign the messages they want to send such as RREQs and RREPs in order to authenticate themselves to the destination nodes. This signature protects the non-mutable information of the AODV messages, which is all the information apart from the hop count field that changes in every transmission of the message in a hop-by-hop frequency until it reaches the destination. Likewise, SAODV uses another scheme to protect the hop count information based on the concept of hash chains by using message digests mechanisms. The protocol uses asymmetric cipher and each node has to store a pair of keys and the authenticated public keys of the other nodes. However, SAODV is considered enough strong to defend MANET communications against different kind of attacks, the asymmetric cryptographic schemes that uses highlights the protocols inappropriate in terms of energy consumption and speed for lightweight handheld devices. In addition, the authors in [22] show that the IPSec security for AODV is in general more efficient in terms of total routing load, throughput and packet delivery fraction than SAODV. At the same time the utilisation of IPSec provides MANET with the security features that SAODV offers.

AODV-SEC [23] is an extension of SAODV that uses a PKI as a trust anchor therefore nodes can be identified using certificates. ARAN (Authenticated Routing for Ad-hoc Networks) [24] provides a solution for securing routing in the managed-open environment. The protocol specifically provides authentication and non-repudiation services by using pre-determined cryptographic certificates. The SAR (Security-aware Ad-hoc Routing) [25] protocol assumes a trust hierarchy in a way that nodes lower in the hierarchy are less trusted than nodes belong to the higher levels. This categorisation determines the way of the routing procedure. To this end, nodes do not select the shorter path to a destination but the one that includes the most trusted nodes.

In [6] the authors propose a protocol to secure the DSR (Dynamic Source Routing) protocol. The name of this is SRP (Secure Routing Protocol) and it assumes that there is a bidirectional security association (SA) between nodes that desire to exchange messages and a shared secret key. The latter is used to sign the non-mutable fields of the routing messages to ensure integrity. They also claim that by the aforementioned signature DSR provides sufficient security for the routing procedure. An alternative solution to SRP is Ariadne [7], which provides authenticity of information provided by the intermediate nodes on the path between a source and a destination in addition to the features that SRP offers. Ariadne is actually similar to SAODV allowing nodes to authenticate routing messages and verify the integrity of them. However, Ariadne is more complicated than SAODV due to RREQ messages in DSR being modified by each forwarding node to include their own address. In [26] a scheme has been proposed for securing OLSR by using digital signatures for hop-by-hop or end-to-end authentication.

The majority of the aforementioned solutions are implemented using asymmetric cryptographic keys. According to [8] asymmetric cryptography is 1000 times slower than symmetric in addition to the fact that for low power devices such as PDAs the battery consumption is too high when the former is used. Furthermore, most of these works secure only one specific protocol giving less flexibility in cases where we want to utilise an adaptive MANET routing protocol. Consequently a unified security mechanism has to be implemented to support secure routing for MANETs.

## 3. ROUTING FOR eMANETs

Within the context of eMANET multimedia communications operating within a pre-defined disaster area (referred in this article as the *Critical Area*), we have designed and developed a novel hybrid and adaptive routing protocol called ChaMeLeon (CML). The protocol is a work in progress [1] within the realm of IETF. The main concept behind CML is the adaptability of the utilized routing mechanism towards changes in the physical and logical state of the network so that the overall performance of the routing algorithm is improved. The importance of such an approach resides in the fact that the nodes in eMANETs have to provide a certain level of routing Quality of Service (QoS) to support multimedia communications but concurrently to cope with limited resources.

In the case of extreme emergency operations, the number of rescuers within the Critical Area (CA) is likely to vary whenever rescuers join or leave the operations according to the severity of the situation. In the case where an eMANET is deployed, the total number of nodes in the network will vary as the number of participating devices join or leave the network. In addition, the battery exhaustion of lightweight communicating devices used by rescuers might stipulate another reason for changes in eMANET sizes. Hence, the eMANET nodes have to be able to efficiently and effectively route data packets considering such extreme scenarios. Thus, for small networks, CML routes data proactively using OLSR protocol [4] whereas for larger networks it uses the reactive AODV protocol [2] mechanisms so that the overall routing performance is improved as supported by our results. It is also important to note in the sphere of multimedia communications supported by eMANETs, the routing protocol efficiency can be quantitatively defined using routing QoS metrics such as delay and jitter (delay variations). These will also be used in this article to define the efficiency of discussed protocols.

The OLSR protocol and the reduced flooding MPR concept are described in [4]. It can be deduced that it readily provides a periodically updated route for data transmission. On the other hand, AODV protocol is explained in [2] and the on-demand route establishment process adds delay towards end-to-end data delivery. However, this allows AODV to provide a fresh route for data delivery as opposed to periodically pre-established routes. It can be deduced from these descriptions that average end-to-end packet delivery delay is reduced in OLSR because of pre-established routes. However, in dynamic and mobile networks comprising of large number of nodes, routes from source to destination nodes can change rapidly. In such cases, packets will be lost and messages would have to be retransmitted adding to more congestion and packet delivery delay. For a reactive approach such as AODV, routes are freshly established on-demand and therefore results in a more efficient approach for large dynamic networks. In addition, the increased size of the periodically exchanged TC messages coupled with the higher required number of such messages in large networks implies that AODV is a better alternative than OLSR for larger networks. For smaller networks, OLSR imposes less delay on data delivery due to pre-defined routes while the fresh routes of AODV provide little advantage as compared to the added route discovery delay in a small network where nodes are within close proximity and where route changes are less acute. These observations are supported by simulation results provided later in this article.

Due to the design of CML, it is important to define a Network Size Threshold (NST) that differentiates a small network from a large network, i.e. the network size beyond which AODV becomes more efficient than OLSR. After a series of simulations where end-to-end packet delivery delay was used to estimate the efficiency of the network, results showed that beyond 10 nodes, AODV outperforms OLSR. In addition, the packet jitter statistics confirm that OLSR is more efficient for networks containing 10 nodes or less, whereas for larger networks, AODV performs better than OLSR. Thus, for the purpose of this article the NST value is set to 10 nodes. In this article, the simulation results are depicted in the section of *performance evaluation*. The protocol consists of 3 phases of operation, which are the reactive phase (*r-phase*), the proactive phase (*p-phase*) and the oscillation phase (*o-phase*). In the p-phase, which is also the default operating phase, CML uses the OLSR proactive approach for routing while in the r-phase, it routes data packets reactively as described by the AODV routing protocol. The r-phase and p-phase are also referred to as the stable phases of CML operation. Moreover, CML introduces an *Adaptive Module*, which is responsible for monitoring the total number of nodes in the network and for acting accordingly if the NST has been exceeded. There are two possible routing scenarios, which can result in such an event.

Firstly, in small networks or at network set up time, CML operates in p-phase. Each time a control TC message is received, the Adaptive Module of a node checks for the number of nodes in the network. This task consists of computing the total number of accessible nodes from the routing table, which is populated proactively. If the number of nodes exceeds the NST, the Adaptive Module has to shift node routing operation to the o-phase. Then, for large networks, CML operates in the r-phase. Each time a RREP is received at a source node, the Adaptive module checks for the number of nodes in the network. Since AODV does not store information of all reachable nodes in the network, the hop count value from the RREP packet is used to estimate for the total number of eMANET nodes. Then, assuming that the nodes in the network are uniformly distributed, the total number of nodes can be estimated as being directly proportional to the square value of the maximum hop count of the network. Hence, the Adaptive Module has to shift node operation to the o-phase only if the RREP indicates that the network size is less than or equal to the NST value.

The Monitoring Module is however unable to detect the events of node oscillations. Oscillation takes place when a node or a group of nodes frequently joins and leaves the eMANET so that the total number of nodes varies beyond and back below the NST repeatedly. Therefore, oscillations would cause the nodes to constantly change routing phases and would have been highly inefficient. Instead, the o-phase is responsible for confirming whether the Adaptive Module has indeed detected an actual change in network size or whether an oscillation has taken place while at the same time allowing the current stable phase of operation to continue. Since oscillation is characterized from the number of oscillating nodes and the oscillation time period, the solution proposed by the o-phase is twofold.

In the first place, the o-phase confirms that the NST value of the network has been reached. Nevertheless, the NST value is now considered as "10+x", as a solution for group oscillations if the p-phase is operating. If the node is running in the r-phase, NST is equal to 10-x. Here, the value of "x" is equal to the tolerance level for group oscillations and indicates that the eMANET should not be sensitive to network changes caused by x nodes oscillating together. The o-phase confirmation process also differs depending on the stable operation phase. If the current routing phase is the p-phase, the o-phase checks the number of nodes twice more whenever a TC message is received. If at least one of the checks confirms that the NST value has been reached, the o-phase shifts to r-phase operation. Otherwise, the o-phase resumes pure p-phase operation.

In the case that the node is routing in r-phase but operating in o-phase, the source node floods a Hop Count Request packet (HCREQ) in the eMANET, containing a TTL value of Network Hop Threshold (NHT). As described previously, NHT is estimated using the NST value and the relationship between the maximum hop count and the total number of nodes in the network. If a node receives a HCREQ packet and its TTL value is zero, the node has to send a unicast Hop Count Reply (HCREP) packet to the source. If the TTL value is more than zero, the TTL value is decreased by one and flooded back into the network. Additionally, in such a case, the node has to generate and flood its own HCREQ termed its echo-HCREQ. As a result, if it receives an echo- HCREP, the node forwards this to the source node of the HCREQ, which initiated the corresponding echo-HCREQ. The source node has to wait for a HCREP or echo-HCREP for a time period of "4*network traversal time" as defined in [4] each time it issues a HCREQ. This is to make sure that the network has been fully probed to verify if the maximum achievable hop count in the network is indeed less than NHT. The source node has to send an HCREQ in two occasions as part of the o-phase confirmation process. If no HCREP or echo-HCREP is received at least once during the confirmation, the o-phase determines that the network size is below the NST and it shifts to p-phase operation. If not, the r-phase is resumed.

The o-phase also has the responsibility to alert the remaining nodes in the network if a shift in routing phase occurs within the node. It floods the eMANET with a Change Phase (CP) packet containing information about such a phase shift. Any node that receives a CP packet forwards it to its Adaptive Module which checks whether the current stable operating phase is the same as the phase to which the CP packet requires the node to shift to. If this condition is true, no further action is undertaken. Otherwise, the Adaptive Module shifts routing operations to the o-phase, which in turn shifts operation to the phase indicated in the CP packet. In all the aforementioned cases, the CP packet is flooded back into the eMANET. Furthermore, the o-phase maintains an oscillation timer, which prevents ineffective phase shifts due to periodic node oscillations. The timer is set to the value of $T_{osc}$ each time a phase shift occurs. If the timer is still valid, the Adaptive Module cannot initiate a switch to the o-phase.

Other popular protocols that propose a scalable routing mechanism for varying network sizes include Zone Routing Protocol (ZRP) [11], Landmark Ad Hoc Routing Protocol (LANMAR) [12] and Sharp Hybrid Adaptive Routing Protocol (SHARP) [13]. The disadvantages of using ZRP reside in the fact that firstly, it does not use the well established AODV and OLSR protocols for routing. Then, even in large networks, the intra zone routing mechanism still requires that nodes periodically send route discovery and maintenance messages on top of any inter zone route discovery messages sent, resulting in more overhead and delay than the CML protocol for such a scenario.

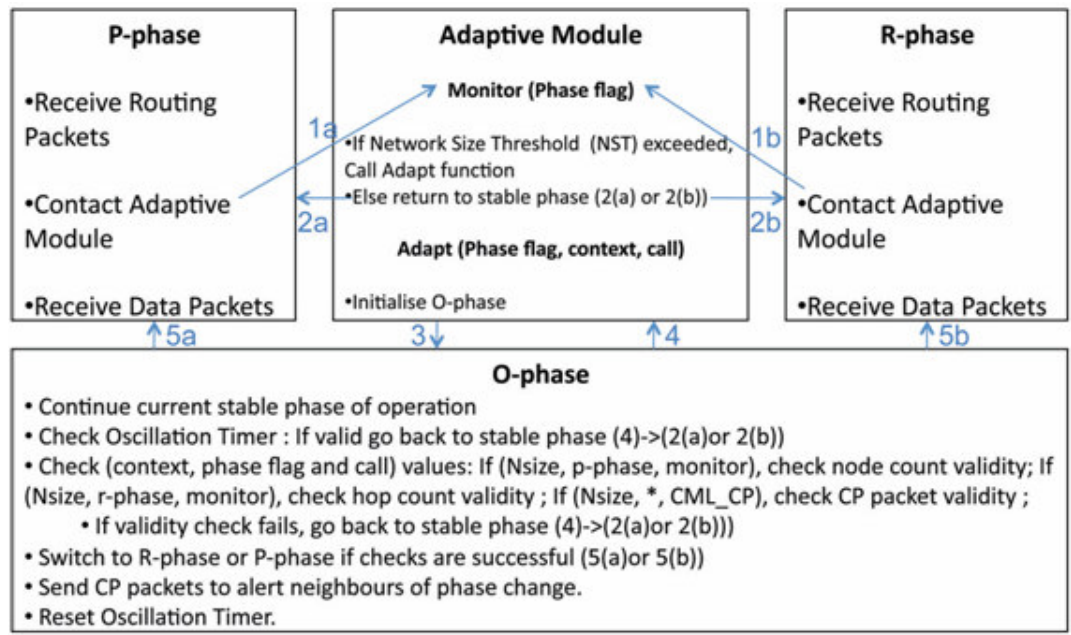

Figure 3. Overview of the CML protocol [1].

In the case of eMANETs different group of nodes are expected to merge together and partition from each other regularly as part of rescue missions. This will cause LANMAR nodes to regularly elect new landmarks as well recalculating its tables frequently to include or exclude landmarks and storing new subnet information. Such a process will add more delay and overhead as compared to CML. Finally, instead of using the number of nodes to shift to a homogenised AODV or OLSR protocol operation as in CML, SHARP proposes to vary the OLSR radius (number of hops) coverage of each node and route data using AODV for nodes outside the radius similar to the way ZRP operates. However, although SHARP guarantees delay jitter performance improvements as compared to AODV, it cannot guarantee improvements in end-to-end packet delivery latency as compared to AODV, OLSR and consequently CML. In figure 3 we have summarised the functionality of the CML protocol.

It is worth noting that CML has the potential to change its routing functionality based on different constraints and requirements along with the network size. However, this multidimensional routing solution must be designed carefully in the future, to prevent NP-completeness.

## 4. SECURITY CONSIDERATIONS

Security considerations about MANET routing protocols are critical because of the absence of a fixed infrastructure. Nodes are associated and will cooperate virtually with any node including adversaries. Adversaries can cause the disruption of the route discovery and data forwarding operations. Disruption of the route discovery mechanism can cause systematic problems to the flow of data. Adversaries can obstruct the propagation of legitimate queries and routing updates. In order to prevent such attacks it is important for the receiver node to verify the authenticity of the sender and the integrity of the data. In addition, malicious nodes can overhear the wireless channel and learn confidential and private information about legitimate nodes.

### 4.1 CML Vulnerabilities

According to [1], CML does not specify any special security countermeasures against malicious entities. However, many secure versions of AODV and OLSR have been proposed in bibliography CML introduces new vulnerabilities. For instance an adversary can send a change phase packet to call the o-phase of CML and the routing behaviour to change accordingly. In this way, CML will not operate in the proper routing mode and the MANET's performance will not be optimal considering the real number of nodes in the network. In addition, legitimate nodes will flood the network with the CML CP packet and will generate traffic overhead. In another case, a set of malicious nodes that coordinate their actions against the CML routing functionality may periodically come into and depart from the network in order to frustrate the cognitive capability of the CML. In this context, CML recognizes that the number of nodes in the MANET has changed and oscillates from the proactive phase to the reactive or vice-versa. The continuous oscillation of CML can result in draining the battery level of the emergency devices very rapidly. Another version of the above attack is launched when malicious nodes change the "hop value" in the CML HCReq packet. In this case, legitimate nodes believe that the size of the network has changed and CML oscillates unreasonably. On the other hand, if a phase shift should take place (due to a real change of the number of nodes) but malicious nodes succeed to drop some or all of the Change Phase packets, the performance of the MANET will not be optimized. According to [1], security effective approaches have to be integrated into CML to avoid the aforementioned attacks and to provide eMANET nodes with basic security requirements such as *confidentiality*, *authentication*, *integrity* and *availability*.

### 4.2 IPSec over CML

Taking advantage of the strength of Security Architecture for the Internet Protocol (IPSec) [13] we have secured the CML routing protocol subsequently creating the Secure CML (SCML) protocol. SCML uses a hybrid version of the IPSec protocol, which includes both AH and ESP modes [16]. This choice is done based on the fact that a potential attacker can perform *traffic analysis*, examine protocol numbers, modify the destination address, and other IP fields if the only applied protocol is ESP. This happens due to the fact that the ESP does not protect the IP header of an IP datagram. By merging ESP with AH the aforementioned harmful situations are prevented efficiently. Additionally, many researchers have argued in favour of a more compact and efficient version of IPSec which will not include the AH. Although, in transport mode, ESP provides integrity protection the latter is not the same with the one provided by AH. ESP offers integrity protection for everything beyond the IP header; AH provides integrity protection for some of the fields of the IP header. Research also suggests that as far as ESP in tunnel mode has the potential to provide the same degree of authentication of the original IP header and integrity of payload with AH, it is not reasonable for both of them to exist. However, in our work we use the transport mode, as this is far more suitable than tunnel mode [16] for MANETs. The benefit of this choice is the avoidance of useless and intolerant overhead. We illustrate the different modes of the IPSec protocol in Figure 3.

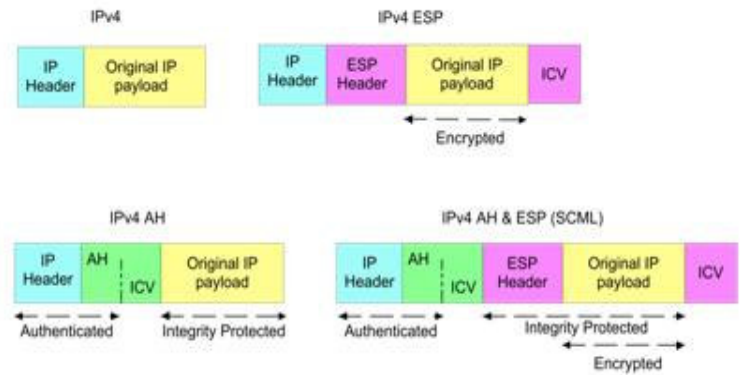

Figure 3. Different IPSec setups and the case of SCML.

ESP offers confidentiality by encrypting the IP payload using 128-bit symmetric *Advanced Encryption Standard* (AES) keys and AH offers authentication and integrity of transmitted packets. The *Rijndael algorithm* has been selected as the algorithm of choice for the AES. It is a symmetric block cipher that supports different key and block sizes of 128, 192 and 256 bits with the AES standardized to be the fixed block size of 128 bits. The most important characteristic of the algorithm is the fact that it combines implementation convenience and simplicity with increased protection against different attacks. Different encryption algorithms could be used in conjunction with IPSec such as the Data Encryption Standard (DES), 3DES, AES, Rivest Cipher 5 (RC5), International Data Encryption Algorithm (IDEA), 3-IDEA, CAST[1], Blowfish and others. However we have used the AES algorithm due to its proven strength and its low overhead.

For authentication and integrity, nodes run the keyed-*Hash Message Authentication Code - Message Digest 5* (HMAC-MD5) algorithm. HMAC is a type of Message Authentication Code (MAC) calculated using a cryptographic hash function in conjunction with a secret key. On the other hand, MD5 is a one-way hash function, which processes input text in 512 bit blocks to generate a 128-bit hash value. The hash values are used afterwards for the verification of the sent message. It is worth mentioning that although MD5 has been found vulnerable to some attacks, when it is used in conjunction with HMAC it is not compromised. Another cryptographic hash function is the *Secure Hash Algorithm 1* (SHA-1), which processes input text in 512 bit blocks to generate a 160-bit hash value.

An apparent question might be "*why are we benefited by using IPSec*?" IPSec is a protocol suite for securing IP-based communications focusing on authentication, integrity[2] confidentiality and supports perfect security forward [14]. The significant importance of the aforementioned protocol is that it offers flexibility, which cannot be achieved at higher or lower layer abstractions in addition to the symmetric cryptographic schemes. These are 1000 faster than asymmetric cryptographic schemes, a fact that makes IPSec appropriate to be used in handheld resource constrained devices such as PDAs. In this context, several research approaches such as [15] and [16] have concluded that the usage of IPSec is appropriate in MANETs. It is widely accepted that IPSec is one of the best security protocols available at present and it is mentioned as the most reliable and efficient network layer protocol. An intruder who eavesdrops on a wireless communication link has the potential to capture passwords which are being transmitted unencrypted. The malicious node can then use the said password to masquerade as a legitimate mobile user. Using the IPSec protocol, data packets are encrypted and the attacker cannot capture passwords or overhear private information. Additionally, an intruder can spoof his IP address to masquerade as a trusted node when the address-based authentication scheme is used. IPSec protects against IP spoofing attacks by deploying strong authentication techniques. Denial of Service (DoS) attacks furthermore could furthermore be avoided by using IPSec. Specifically, when an intruder is trying for instance to launch a Transmission Control Protocol (TCP) SYNchronous (SYN) flooding attack by sending a sequence of connection request messages, the available buffer space of the target-victim system is overrun. Due to the strong authentication that IPSec offers, the intruder launches TCP SYN flooding attacks using its own IP address, revealing its location and identity. IPSec encryption techniques moreover protect the network from a potential session hijacking by an intruder. The latter does not know the encryption keys required to decrypt and encrypt the data stream. Finally, in keeping with the concept of integrity protection that IPSec provides, when the Integrity Check Value (ICV) of a packet is valid it receives the appropriate treatment and the nodes decide the next hop node across the path to the destination. If any unauthenticated node changes any data in the IP datagram or updates the ICV, this node will be detected and the packet will be discarded.

---

[1] It takes its name from its creators namely Carlisle Adams Stafford Tavares.
[2] Includes the replay attack prevention.

## 5. PERFORMANCE EVALUATION

Authors in [17] discuss the space[3] and time complexity[4], introduced by the different modes of IPSec deployment, in terms of number of CPU cycles. In the same work, they find out that the total number of operations required for MD5 processing per 512 bits block is 720 plus 24 operations for initialization and termination, while for SHA-1 processing is 900 plus 210 operations for initialization and termination. Consequently, the overhead introduced by HMAC-MD5 is lower. In order to compute the exact time of HMAC-MD5 operation for an input of packets $n_k$ and for processor speed $c_p$ the following equation is used:

$$t_{\text{HMAC-MD5}}(n_k, c_p) = [32 + (2 + 744\ n_k)] / c_p \quad (1)$$

To go a step further, authors in [17] derive the corresponding number of processing cycles required for encrypting one block of data with each one of the three standardized types of AES for different key lengths 6168, 7512 and 8856 number of CPU cycles for 128, 192 and 256 key length respectively. For the algorithms weaker than AES namely DES and 3DES, the corresponding overhead is 2697 and 8091 CPU cycles. The choice of the AES algorithm[5] to generate the symmetric keys, which will be used by the ESP to encrypt the payload of the IP datagram has been done based on the fact that the algorithm is the fastest and cryptographically strongest. The time overhead of AES is:

$$T_{\text{encryption}} = 6,168 \text{ and } T_{\text{decryption}} = 10,992 \text{ processing cycles per packet (2)}$$

IPSec packetisation and ciphering increase the size of the transmitted packets. In the transport mode, the space overhead of AH is equal to 24 and ESP equal to 10 bytes [18]. The size overhead of ESP is 10 bytes considering no authentication. Consequently, the space overhead in the case of SCML[6] is (24+10) = 34 bytes. In order to simulate the case of an eMANET we use the Mission Critical Mobility (MCM) model developed for the purposes of the PEACE project and published in [19]. MCM implements the two-way ground propagation model and the Random Waypoint mobility model considering obstacles. MCM is a mobility model that captures the properties of the mobility of the nodes (firemen, policemen, paramedics, etc.) of eMANETs. We have supposed that the emergency workers are equipped with PDAs with processing capability equal to 450 Millions of Instructions Per Second[7] (MIPS) in order to compute the delay introduced in each device of HMAC-MD5 and AES encryption or decryption algorithms. From the equations (1) and 2, we have derived the time overhead per packet for each of HMAC-MD5 and AES algorithms. Namely:

$$t_{HMAC-MD5} = 1.68 \text{ μsec/ packet}$$

$$t_{AES,encryption} = 13.7 \text{ μsec/ packet}$$

$$t_{AES,decryption} = 24.4 \text{ μsec/ packet}$$

---

[3] Security related additional information on the transmitted packets increases the bandwidth consumption.
[4] Security processing increases the packet end-to-end delay and the transmission time.
[5] With 128 key length

The comparative simulation results for CML, AODV, OLSR and DSR for varying network sizes from 5 to 50 nodes are shown in Figures 4-7. The comparison metrics that have been chosen are *average data end-to-end packet delay* and *cumulative delivery delay*, *cumulative packet jitter* (which are crucial for multimedia streaming in eMANETs) and *control routing load* of each protocol. The simulation results highlighted in Figure 4 show that DSR has the worst performance compared to the other routing protocols for the greatest amount of network sizes. We notice that AODV is better than OLSR for network sizes larger than 10 nodes. CML outperforms the rest of the routing protocols in terms of average packet end-to-end delay according to Figure 4.

In Figure 5 we have depicted the cumulative end-to-end delay for the different routing protocols. This illustrates the overall delay of a network that increases its size from 5 to 50 nodes. It is worth noting that the graph of OLSR is always under the AODV. This can be explained as discussed before that OLSR introduces significant less delay for smaller network sizes (smaller or equal than 10) than AODV. This fact, keeps the curve of OLSR under the AODV but the degradation is higher in OLSR for large networks and we can confirm this from the gradients of the two curves. The gradient of OSLR curve is higher than the AODV one.

Figure 6 describes the cumulative average packet jitter performance of the discussed protocols. It can be deduced from the figure that the cumulative jitter performance of the DSR protocol is the worst compared to other protocols over the whole range of investigated network sizes. Also, the OLSR protocol has better jitter performance for small networks with a size of less than or equal to 10 nodes whereas for larger networks greater than 10 nodes the AODV provides better jitter performance. However, CML routing has slightly less jitter performance for the whole range of investigated network sizes as compared to all the other protocols.

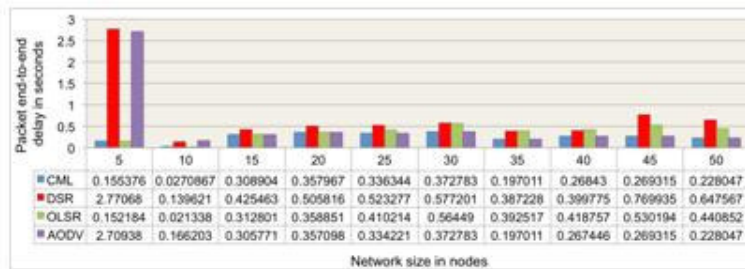

Figure 4. Packet end-to-end delay for each routing protocol against network size.

---

[6] SCML adopts a hybrid solution of AH and ESP.
[7] 450 MIPS is a realistic value for a well-known PDA.

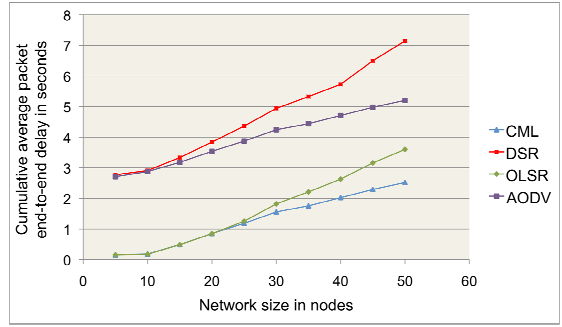

Figure 5. Average cumulative packet end-to-end delay for each routing protocol against network size.

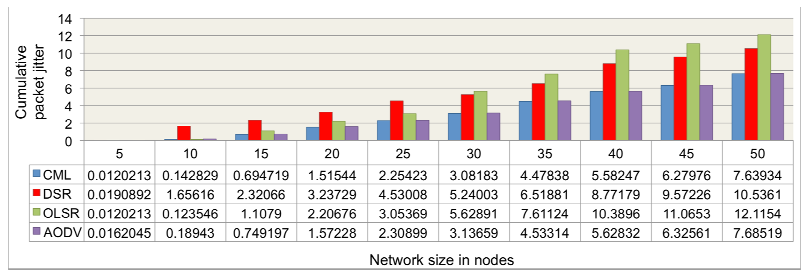

Figure 6. Cumulative packet jitter for each routing protocol against network size.

The total routing load statistics for each protocol are shown in Figure 7. The results show that the routing protocols require more routing control packets as the number of nodes in the network increases. Moreover, OLSR utilises the least routing load amongst all the compared protocols in the investigated scenarios. Overall, DSR routing requires more routing load than the other protocols. Then, AODV uses less routing load than OLSR for small networks of less than or equal to 10 nodes. For larger networks, it emits more control load than OLSR. CML utilises control messages of OLSR for networks smaller than or equal to 10 nodes and utilizes routing messages of AODV for larger networks.

Therefore, the control load used by CML is slightly greater than that of OLSR for smaller networks but at the same time is more than that used by AODV. For larger networks, the routing load of CML is slightly greater than that of AODV but is also greater than routing load of OLSR. The slightly greater routing load of CML compared to its current stable phase of operation, can be accounted for by the CML specific control packets such as the CP, HCREQ and HCREP packets. This can be further confirmed by the fact that for a scenario with an eMANET of 5 nodes, the OLSR and CML routing loads are equal because the CML specific packets are not required for such network sizes.

These results clearly demonstrate that CML takes advantage of the AODV and OLSR routing mechanisms for the different network sizes. CML improves routing delay and jitter efficiency of as compared to AODV, OLSR and apparently to DSR. The aforementioned metrics are considered critical within the context of emergency MANETs where QoS is very important for cohesive communications.

In Figures 8-10, we illustrate the *cumulative packet end-to-end delay*, *the cumulative control load overhead* along with the *cumulative goodput* (is equal to the ratio *data packets/ control packets*) for the different IPSec modes and for SCML. As we expected, Figure 8 shows that SCML introduces the highest overhead all of them because it adopts both AH and ESP modes.

However, the delay overhead of SCML is very near to the scenario of applying ESP mode, which means that the addition of AH mode does not introduce significant overhead. On the other hand, the security benefits, which have thoroughly been discussed in the previous section, are significant. In addition, the simulation results confirm that encryption and decryption, using AES (ESP mode) is slower than HMAC-MD5 (AH mode) and that is the reason of the obvious difference in the delay values of the two protocols.

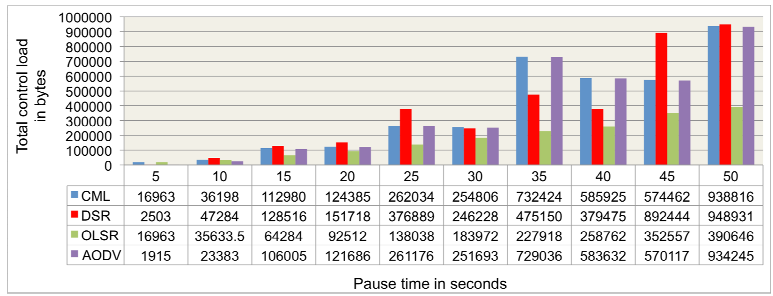

| | 5 | 10 | 15 | 20 | 25 | 30 | 35 | 40 | 45 | 50 |
|---|---|---|---|---|---|---|---|---|---|---|
| CML | 16963 | 36198 | 112980 | 124385 | 262034 | 254806 | 732424 | 585925 | 574462 | 938816 |
| DSR | 2503 | 47284 | 128516 | 151718 | 376889 | 246228 | 475150 | 379475 | 892444 | 948931 |
| OLSR | 16963 | 35633.5 | 64284 | 92512 | 138038 | 183972 | 227918 | 258762 | 352557 | 390646 |
| AODV | 1915 | 23383 | 106005 | 121686 | 261176 | 251693 | 729036 | 583632 | 570117 | 934245 |

Figure 7. Total routing load for each routing protocol against network size.

In Figure 9, we illustrate the performance of each case in terms of total control[8] data. We observe that SCML has the highest value of control data in bytes. This is an acceptable result due to the fact that each SCML packet has an overhead of 192 bits for HMAC-MD5 and 80 bits for AES encryption, a total of 272 bits, more than conventional CML. In terms of data load versus control data load in bytes, the results illustrated in Figure 10 validate the intuition that the ratio is the highest for CML due to the absence of extra security overhead introduced by the rest of the protocols.

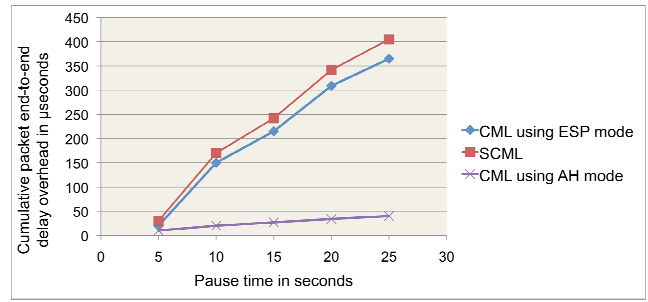

Figure 8. Cumulative packet end-to-end delay overhead.

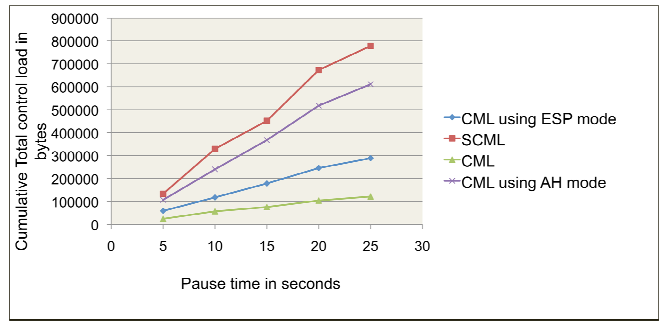

Figure 9. Cumulative control load in bytes for different IPSec modes in CML.

---

[8] Or else routing data.

Generally speaking, we observe that SCML does not introduce unaffordable overhead to CML. Thus, SCML has two main characteristics namely: (i) it is a secure routing protocol according to the arguments mentioned in previous sections (ii) it is based on the efficient CML routing protocol. This is proof enough to highlight the suitability of the protocol when taking into account eMANETs where QoS efficient routing solutions must be adopted.

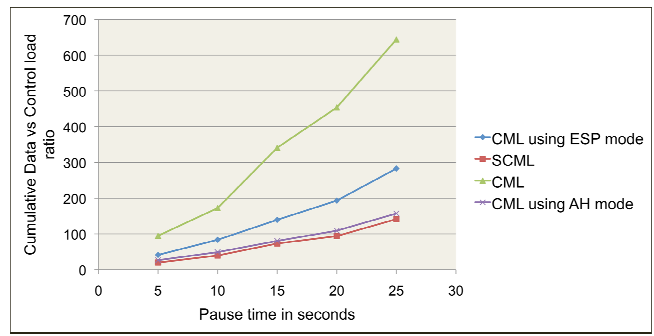

Figure 10. Cumulative ratio data against control load.

## 6. CONCLUSIONS

Mobile ad-hoc networks (MANETs) can be utilised in case telecommunication infrastructures have been collapsed during extreme emergencies such as forest fires, terrorist attacks, earthquakes etc. These networks can replace current technologies such as TErrestrial Trunked RAdio (TETRA) to provide benefits such as interoperability, video services, secure communications for emergency communication with the rescue field.

Due to the decentralised nature of MANETs, nodes must forward packets towards a destination by following the principles of a MANET routing protocol. For the purposes of emergency MANET routing we propose in this article an adaptive routing protocol called ChaMeLeon (CML) [1], supporting emergency communications. CML is a hybrid and adaptive routing protocol operating within a defined disaster area denoted as the *Critical Area*. The main concept behind CML is the adaptability of its routing mechanisms towards changes in the physical and logical state of a MANET. We have proven that CML outperforms AODV, OLSR and DSR in terms of *average end-to-end data packet delay, cumulative delivery delay* and *cumulative packet jitter* (which are crucial for multimedia streaming in eMANETs) while it has slightly increased *control routing load*. Additionally, we applied IPSec on top of CML to provide *confidentiality*, *authentication* and *integrity* to the transmitted packets. These are essential requirements in the presence of malicious nodes that try to disrupt the conventional operation of CML.

We additionally secure this protocol by taking advance of the IPSec protocol. Specifically, we use both modes of the IPSec protocol namely AH and ESP to develop a hybrid IPSec mode in transport mode. This mode is appropriate to provide the aforementioned security requirements for CML. The performance evaluation shows that the choice of hybridization does not introduce unaffordable overhead to SCML.

We compared the performance of CML against well-known routing protocols OLSR, AODV and DSR in terms of average end-to-end packet delay and jitter along with the cost of routing load. Simulation results show that CML is in overall the most efficient solution to be adopted for emergency MANETs where the network size frequently changes during a rescue mission. Furthermore, we noticed that the time and space overhead introduced by the use of IPSec is affordable whilst at the same time IPSec guarantees confidentiality, authentication and integrity to CML.

# 7. FUTURE WORK

Our plans for future work include but are not limited to guarantying security requirements, assuming that even if malicious nodes succeed to capture a node (node-capture attack), intrusion detection mechanisms such as the one proposed in [20] can be applied as a second wall of defence. In addition, we are planning to integrate the upcoming version of OLSR entitled OLSRv2 along with the DYnamic Manet On-demand routing protocol (DYMO) and to compare the new version of CML with the non-adaptive protocols.

## ACKNOWLEDGEMENT

This work was undertaken in the context of the project ICT SEC-2007 PEACE (IP-based Emergency Applications and serviCes for nExt generation networks) with contract number 225654. The project has received research funding from the European 7[th] Framework Programme.

## Authors

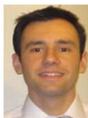

**Emmanouil A. Panaousis** is currently a research Ph.D. student at Kingston University, UK, Faculty of Computing, Information Systems and Mathematics (CISM). He works within a team on Wireless Multimedia Networking (WMN) Research Group. Emmanouil received his M.Sc. in Computer Science with distinction at the Department of Informatics of the Athens University of Economics and Business and his B.Sc. in Informatics and Telecommunications at the National and Kapodistrian University of Athens. Emmanouil has published more than 15 papers in international journals and conferences and one book chapter. Emmanouil is a student member of the British Computer Society, the IEEE and the IEEE Communications Society. E-Mail: e.panaousis@kingston.ac.uk

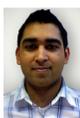

**Tipu Arvind Ramrekha** is a PhD student at Kingston University, UK, Faculty of Computing, Information Systems and Mathematics (CISM). There he is part of a research team on Wireless Multimedia Networking (WMN) and currently researching on, "Quality of Service routing for multimedia communications in Mobile Ad-hoc Networks for extreme emergency cases", in the CISM faculty. He holds a first class B. Eng. in Computer Engineering from NSIT, Delhi University, India. He then pursued successfully his MSc in Networking and Data Communications with management studies at Kingston University, London, UK where he received his MSc with a distinction and was awarded the best CISM postgraduate student and project awards. He is a member of the IEEE. E-Mail: a.ramrekha@kingston.ac.uk

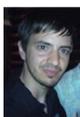

**Grant P. Millar** is currently a Ph.D student at the Faculty of Computing, Information Systems and Mathematics (CISM), Kingston University, London, UK. He works within a team on Wireless Multimedia Networking (WMN) as well as giving the occasional lecture on various subjects within the networking realm within the CISM faculty. Grant received his M.Sc in Networking and Data Communications at the faculty of CISM, Kingston University and his B.Sc in Computer Animation at the University of Portsmouth, UK. Grant is a student member of the IEEE and the IEEE Communications Society. E-Mail: g.millar@kingston.ac.uk

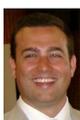

**Christos Politis** is a Senior Lecturer (Ass. Prof) at Kingston University London, UK, Faculty of Computing, Information Systems and Mathematics (CISM); where he leads a research group on Wireless Multimedia & Networking (WMN) and teaches modules on wireless communications in the CISM faculty. Prior to this he was the Research and Development (R&D) project manager at Ofcom, the UK Regulator and Competition Authority. Christos was for many years he was a post-doc research fellow at the Centre for Communication Systems Research (CCSR) at the University of Surrey, UK. He is/ was involved with several EU (IST and ICT), national and international projects, and was the project manager of the IST UNITE. Christos is a patent holder, and has published more than 70 papers in international journals and conferences and chapters in two books. Christos was born in Athens, Greece and holds a PhD and MSc from the University of Surrey, UK and a B.Eng. from the Technical University of Athens, Greece. He is a member of the IEEE and Technical Chamber of Greece. E-Mail: c.politis@kingston.ac.uk